\documentclass[aps,pra,twocolumn,amssymb,amsfonts,floatfix,superscriptaddress,showpacs]{revtex4}

\usepackage{graphicx}
\usepackage{dcolumn}
\usepackage{bm}
\usepackage{mathbbol}
\newcommand{\ket}[1]{|{#1}\rangle}
\newcommand{\bra}[1]{\langle {#1}|}
\begin{document}

\title{Quantum Chaos, Delocalization, and Entanglement in Disordered
Heisenberg Models}

\author{Winton G. Brown}
\email{Winton.G.Brown@Dartmouth.edu}
\affiliation{\mbox{Department of Physics and Astronomy,
Dartmouth College, 6127 Wilder Laboratory, Hanover, NH 03755, USA}}

\author{Lea F. Santos} \thanks{Current address: Department of Physics,
Yeshiva University, New York, NY 10016, USA. Email: {\tt lsantos2@yu.edu}}
\affiliation{\mbox{Department of Physics and Astronomy, Dartmouth
College, 6127 Wilder Laboratory, Hanover, NH 03755, USA}}

\author{David J. Starling}
\affiliation{\mbox{Department of Physics and Astronomy, University of
Rochester, Rochester, NY 14627, USA}}

\author{Lorenza Viola}
\thanks{Corresponding author. Email: {\tt Lorenza.Viola@Dartmouth.edu}}
\affiliation{\mbox{Department of Physics and Astronomy,
Dartmouth College, 6127 Wilder Laboratory, Hanover, NH 03755, USA}}

\date{\today}

\begin{abstract}
We investigate disordered one- and two-dimensional Heisenberg spin
lattices across a transition from integrability to quantum chaos from
both a statistical many-body and a quantum-information perspective.
Special emphasis is devoted to quantitatively exploring the interplay
between eigenvector statistics, delocalization, and entanglement in
the presence of nontrivial symmetries. The implications of basis
dependence of state delocalization indicators (such as the number of
principal components) is addressed, and a measure of {\em relative
delocalization} is proposed in order to robustly characterize the
onset of chaos in the presence of disorder.  Both standard
multipartite and {\em generalized entanglement} are investigated in a
wide parameter regime by using a family of spin- and fermion- purity
measures, their dependence on delocalization and on energy spectrum
statistics being examined.  A distinctive {\em correlation between
entanglement, delocalization, and integrability} is uncovered, which
may be generic to systems described by the two-body random ensemble
and may point to a new diagnostic tool for quantum chaos.  Analytical
estimates for typical entanglement of random pure states restricted to
a proper subspace of the full Hilbert space are also established and
compared with random matrix theory predictions.

\end{abstract}

\date{\today}
\maketitle

\section{Introduction}

The emergence of non-integrable behavior in quantum mechanics is a
fascinating and widespread phenomenon which is largely responsible for
the ``complexity'' intrinsic to the physical and mathematical
description of interacting many-body quantum systems.  A most striking
consequence is the existence of distinctive {\em quantum-chaotic}
properties for dynamical systems which may lack a clear classical
limit.  The characterization of such quantum chaos signatures has a
long history, pioneered by Wigner in his effort to quantitatively
model complex nuclei ~\cite{Wigner}, and eventually culminating in
statistical approaches to complex quantum systems based on so-called
{\em Random Matrix Theory} (RMT) \cite{Izrailev1990,Mehta,RMT_Guhr}.

Recent years have witnessed a renewed interest toward qualitatively
reassessing and quantitatively exploring many-body quantum complexity
and quantum chaos implications in the light of Quantum Information
Science (QIS) \cite{Nielsenbook,Casatibook}.  On one hand, a deeper
understanding of quantum chaos and its implications is a prerequisite
for identifying potentially harmful consequences as well as beneficial
uses of chaos in information-processing devices: While the possibility
that disorder may destabilize quantum computation through a ``chaotic
melting'' \cite{geo1,geo2} calls for careful hardware design and error
control, chaotic evolutions tend to naturally generate effectively
{\em random} states which are a resource for a variety of QIS
protocols \cite{Hayden}.  On the other hand, QIS provides additional
tools for describing complexity of states and evolutions, which are
proving useful in a variety of settings at the interface with
condensed-matter and statistical physics.  The notion of {\em
entanglement}, in particular -- as capturing distinctively quantum
correlations which admit no local classical interpretation
\cite{Schroed} -- plays a central role to this end.  Among the most
notable developments to date, entanglement theory has allowed to
considerably deepen the understanding of quantum critical phenomena --
appropriate entanglement measures serving as an ``order parameter''
for detecting and classifying quantum phase transitions in matter
\cite{Fazio07} -- and to devise enhanced computational methods for
both static and time-dependent properties of quantum lattice systems
\cite{Comp}.

In this context, uncovering the relationship between various aspects
of complexity in quantum states and evolutions, quantum chaos, and
entanglement is both a natural and fundamental challenge, which is
spurring significant activity in the field, see e.g.
\cite{Miller,Prosen,Caves,Jacq,Ghose,Wang,Bandyopadhyay2002,Bandyopadhyay2004,Lakshminarayan2005,Yaakov2005,Montangero2006}
for representative contributions.  In particular, specific questions
to be answered include the following: What entanglement properties
best capture the {\em structural change} occurring in typical
many-body eigenstates across a transition to quantum chaos, and how
well do such properties reflect the complexity of chaotic eigenstates?
To what extent does the amount of entanglement relate to the amount of
underlying {\em state delocalization}?  Perhaps most importantly, can
entanglement theory suggest {\em new signatures} of quantum chaos?

Our goal in this work is to take a step toward answering some of the
above questions, by seeking an in-depth characterization of
entanglement properties of the {\em stationary states} (eigenvectors)
of non-integrable as opposed to integrable many-body Hamiltonians, in
relation to the behavior of traditional complexity indicators related
to RMT spectral statistics and delocalization measures.  In this
respect, our analysis shares some motivation with earlier studies of
entanglement across a transition to non-integrability in a class of
one-dimensional Harper Hamiltonians by Lakshminarayan {\em et al.}
\cite{Lakshminarayan2003}, and spin-$1/2$ lattice systems by Santos
{\em et al.} \cite{lea3}, and Mejia-Monasterio {\em et al.}
\cite{Mejia05} where, however, primary emphasis to {\em pairwise} and
{\em bipartite} entanglement is given.  Only recently has genuine {\em
multipartite} entanglement started to be addressed, notably in the
one-dimensional Ising model with a tilted magnetic field
\cite{Lakshminarayan2005,Lakshminarayan2006}.

Here, we focus on a representative class of {\em disordered Heisenberg
models}, which have received limited attention to date in spite of
their prominent role in condensed-matter physics as well as in
exchange-based circuit-model
\cite{exchange1,exchange2,kane,helium1,Bacon,helium2,SantosPRB} and
cluster-state \cite{Levy} quantum computing architectures.  As a
further distinctive feature of our work, the notion of {\em
Generalized Entanglement} (GE), introduced by Barnum {\em et al.} in
\cite{BKOV03,BKOSV04}, is exploited to both obtain a unified approach
to standard (qubit-based) multipartite entanglement -- quantified by a
family of coarse-grained (spin) purity measures -- and to construct GE
measures directly probing correlations in different (fermionic)
operator languages.  From this point of view, the present study
further validates the usefulness of GE for broadly characterizing
complexity in quantum systems, as recently demonstrated in
applications to ground-state quantum phase transitions \cite{SOBKV04},
chaotic quantum maps \cite{Weinstein}, and efficient solvability of
Lie-algebraic models \cite{Solvab}.

The content of the paper is organized as follows.  We begin in Sec.~II
by recalling the essential RMT background, along with well-established
spectral signatures of quantum chaos and measures of pure-state
delocalization.  Sec.~III introduces the relevant class of
one-dimensional (1D) and two-dimensional (2D) Heisenberg models,
laying out the static disorder settings under examination (associated
with randomness in the one-body energies, the two-body interactions,
or both), and discussing the symmetries associated with different
parameter regimes.  A thorough characterization of integrability
properties is obtained in Sec.~IV, with the twofold objective of
distinguishing between delocalized-localized regions and
chaotic-integrable ones.  The implications of the dependence of
delocalization upon the basis choice are elucidated, and a measure of
{\em relative delocalization} is introduced to gain additional insight
into the properties of eigenvectors in disorder-induced chaotic
regimes.  After a brief account on entanglement and GE measures in
Sec.~V, we present in Sec.~VI a detailed analysis of entanglement as a
function of disorder strength, energy, as well as delocalization
properties.  As a main emerging feature, a strong and persistent
correlation between multipartite entanglement and delocalization is
found in {\em non-integrable} regimes, which is consistent with
independent evidence in \cite{Lakshminarayan2006} and {\em may suggest
a novel signature of quantum chaos}.  While most results are
numerical, analytical estimates for multipartite entanglement of {\em
random pure states localized to a subspace} are also obtained and
compared to RMT predictions. Throughout the paper, special care is
devoted to contrast properties which are general to disordered
spin-$1/2$ lattices to those which are specific to the Heisenberg
interaction.  The paper concludes with a summary and outlook in
Sec.~VII, followed by an Appendix which collects technical
derivations.

\section{Signatures of Quantum Chaos}

It is well established that quantized versions of classically
integrable and fully chaotic systems can be distinguished by their
quantum energy level statistics
\cite{Wigner,Mehta,HaakeBook,RMT_Guhr,Flambaum2,Flambaum1}.  Of
particular interest is the distribution of energy level spacings,
$P(s)$, where $s$ is the spacing between neighboring energy levels
after the spectrum has been appropriately unfolded, so that the
density of states is everywhere equal to one. Integrable systems
typically exhibit a {\em Poisson distribution},
\begin{equation}
P_{\tt P}(s) = \exp(-s), \;\; s \in {\mathbb R}^+,
\label{PD}
\end{equation}
whereas chaotic systems have an energy level spacing distribution
predicted by RMT. Within RMT, an exact description of a complex
physical system is replaced by a statistical description based on
ensembles of random matrices which share the same fundamental symmetry
properties as the original system Hamiltonian.  In particular, for the
wide class of systems exhibiting time reversal invariance, the
appropriate ensemble is the so-called Gaussian Orthogonal Ensemble
(GOE), whose energy level spacing distribution is closely approximated
by the {\em Wigner-Dyson distribution},
\begin{equation}
P_{\tt WD}(s) = \frac{\pi s}{2}\exp\left(- \frac{\pi
s^2}{4}\right),\;\; s \in {\mathbb R}^+
\label{WD}.
\end{equation}

The eigenvectors of fully chaotic systems may also be described
statistically using RMT.  If the system of interest obeys
time-reversal symmetry, the eigenvector components tend to follow a Gaussian
distribution, resulting from the invariance of the GOE under arbitrary
orthogonal transformations \cite{HaakeBook}.

It is important to appreciate that systems such as interacting
lattices of spin-$1/2$ particles have no obvious classical limit, thus
the question of whether or not they may exhibit chaos must be posed
and answered from an entirely quantum-mechanical perspective.  Because
the standard defining features of classical chaos (phase-space
ergodicity and exponential divergence of neighboring trajectories)
have no direct meaning for Hamiltonian quantum dynamical systems,
neither does an unambiguously established framework exist for
consistently defining integrability in quantum settings
\cite{Weigert,Enciso}, such an issue is not straightforward and still
largely open.  In the present context, we shall use the term ``quantum
chaos'' in an operational sense, to simply mean the presence of RMT
energy level statistics.

In order to examine the transition from integrability to chaos, the
following picture is employed \cite{HaakeBook,Flambaum0,Mejia05}. Let
$H_0$ be an integrable Hamiltonian, that is, one for which the
eigenvalues and eigenvectors may be determined analytically
\cite{Sklyanin,Solvab}.
Now consider the effect of an integrability-breaking perturbation, $H'$,
so that the total Hamiltonian becomes
\begin{equation}
H = H_0 + \lambda H', \;\;\;\lambda \in {\mathbb R}^+\,.
\label{basic}
\end{equation}

For sufficiently small values of the parameter $\lambda$, the
eigenvalues and eigenvectors of $H$ are adequately described by
perturbation theory.  As $\lambda$ increases to the point where the
interaction strength between states which are directly coupled by $H'$
is equal to their unperturbed energy difference~\cite{geo2,geo1},
perturbation theory breaks down, and a crossover from a Poisson to a
Wigner-Dyson distribution occurs.  In parallel with such a crossover
in energy level statistics, the eigenvectors of $H$ become
increasingly \textit{delocalized} across the eigenstates of $H_0$.  As
$\lambda$ increases further, delocalization typically continues until
the component distribution of the eigenvectors becomes that of GOE
random states.  The exact relationship between the level statistics
and delocalization border has been much studied and remains in general
an open question~\cite{geo2,geo1,Jacquod2002}.

In order to identify parameter ranges where chaos is present, it is
necessary to quantify how accurately the statistical properties of the
eigenvalues and eigenvectors of the Hamiltonian are described by RMT.
Here, we will address this question by examining the energy level and
eigenvector statistics across the {\em full} spectrum, rather than
restricting to a specific spectral region (see also Sec.~IV.C for
additional quantitative discussion of this point).

The extent to which the energy level spacing distribution distribution
interpolates between the Poisson and Wigner-Dyson limits may be
conveniently parametrized by a so-called \textit{Level Statistics
Indicator} (${\tt LSI}$), introduced in \cite{Jacquod1997}:
\begin{equation}
\eta \equiv \frac{\int_0^{s_o}[P(s) - P_{\tt
WD}(s)]ds}{\int_0^{s_o}[P_{\tt P}(s)-P_{\tt WD}(s)]ds}
\label{LSI},
\end{equation}
where $s_0 \approx 0.4729$ is the first intersection point of $P_{\tt
P}(s)$ and $P_{\tt WD}(s)$. ${\tt LSI}$ has the value $\eta = 1$ when $P(s) = P_{\tt P}(s)$, whereas $\eta = 0$ if $P(s) = P_{\tt WD}(s)$.

In order to characterize the degree of eigenvector delocalization, a
convenient measure is the so-called {\em Number of Principal
Components} (${\tt NPC}$) \cite{Izrailev1990,Zelevinsky1996,memo}.
Given a basis $\{\ket{n}\}$ of the system Hilbert space ${\cal H}$,
${\tt NPC}$ of a normalized pure-state vector $\ket{\psi}$, is defined
as follows:
\begin{equation}
\xi (\ket{\psi}) \equiv \Big{[}\sum_n |\langle n
\ket{\psi}|^4\Big{]}^{-1}\label{NPC}.
\end{equation}

Thus, ${\tt NPC}$ estimates the number of basis states relative to which
$\ket{\psi}$ has a significant component. For example, if $\ket{\psi}$
is a uniform superposition of exactly $m$ basis states, then $\xi(\ket{\psi})= m$.  For an $N$-dimensional state vector with a
component distribution pertaining to the GOE, the expected ${\tt NPC}$
is given by $$\xi_{{\rm GOE}}= \frac{N+2}{3},$$ \noindent
where the factor $1/3$ emerges from the Gaussian fluctuations of the
eigenstates, and the additive correction $2$ is due to
normalization~\cite{RMT_Guhr,Berman2001}. Since $N$ is usually large,
the approximation $\xi_{{\rm GOE}} \sim N/3$ is commonly
adopted.  In parallel with the fact that the ${\tt LSI}$ as constructed
takes into account the level statistics across the full spectrum, we
will be primarily interested in the average value of ${\tt NPC}$ across all of
the eigenvectors in a relevant subspace.

\section{Disordered Heisenberg Models}

The Heisenberg model in one- and two-spatial dimensions plays a
paradigmatic role in condensed matter physics and statistical
mechanics as a testbed for exploring quantum magnetism and spin
dynamics in reduced dimensionality~\cite{Tejada}.  Thus, an accurate
characterization of its integrability properties in physical regimes
of interest has both a fundamental and practical significance.  In the
special case of spin-$1/2$ particles, a partial characterization of
the integrability-to-chaos transition has been achieved, based on both
{\em clean} systems where chaos is induced by coupling two different
spin chains \cite{Hsu1993} or by adding next-nearest-neighbor
interactions \cite{Hsu1993,Kudo2004,Rabson2004}, as well as {\em
disordered} systems, where the integrability-breaking term consists of
random magnetic fields applied to all or a subset of spins
\cite{Avishai2002,Santos2004,Deguchi}.  Aside from their relevance to
model real materials, disordered systems offer the added advantage of
providing a natural arena to study the interplay between interaction
and disorder, which remains a most challenging problem in
condensed-matter physics.

Within QIS, the {\em isotropic} Heisenberg spin-$1/2$ model in an
external magnetic field arises naturally in some of the most promising
solid-state proposals for scalable quantum computation, each spin
corresponding, in the simplest settings, to a logical qubit and the
exchange interaction providing the required inter-qubit coupling.
Following the original suggestion by Loss and
DiVincenzo~\cite{exchange1} for coupling electron-spin qubits via {\em
tunable} Heisenberg interactions in semiconductor quantum dots,
schemes for effecting exchange-based universal quantum computation
have been further developed for both electron~\cite{Vrijen} and
donor-atom nuclear spins~\cite{kane}, constructive methods for
universal quantum gate design and efficient readout being identified
in~\cite{Wu03}.  In addition, scalable universal architectures where
{\em always-on} Heisenberg couplings are used in conjunction with
appropriate encodings of a logical qubits into three or more physical
spins have been constructed, offering both substantial implementation
flexibility~\cite{exchange2} and enhanced decoherence
suppression~\cite{Bacon,Yaakov2007}.

Depending on implementation detail, imperfect qubit fabrication and/or
uncontrolled residual spin-spin couplings during storage or gating may
introduce an integrability breaking perturbation $H'$, causing the
prerequisite mapping to well-defined logical qubits to be lost.  While
a number of error control schemes exist in principle to counteract the
effects of $H'$ (notably, dynamical refocusing methods for static
disorder as considered here,see e.g.~\cite{DD}), understanding the
error behavior due to $H'$ remains an important preliminary step.

The Hamiltonian for an $L$-site lattice of spin-$1/2$ particles
coupled by the Heisenberg interaction and subject to a bias field in
the $z$ direction is given by:
\begin{equation}
H=\sum_{i=1}^{L}\frac{\varepsilon_i}{2}\sigma_z^{(i)}
+\sum_{\{i,j\}}\frac{J_{ij}}{4}\vec{\sigma}^{(i)} \cdot\vec{\sigma}^{(j)}\,,
\label{Ham}
\end{equation}
where $\vec{\sigma}^{(i)}$ is the vector of Pauli matrices
$(\sigma_x^{(i)}, \sigma_y^{(i)},\sigma_z^{(i)})$ acting on the
two-dimensional Hilbert space of the $i$-th site.  The parameter
$\varepsilon_i$ determines the on-site Zeeman energy of the $i$-th
spin, and will be parameterized as $\varepsilon_i = \varepsilon +
\delta \varepsilon_i $, where $\varepsilon$ is a non-zero average and
$\delta \varepsilon_i $ are uniformly distributed within $[-\delta
\varepsilon/2,\delta \varepsilon/2]$, characterizing the different
strengths of local random magnetic fields.  The interaction strength
between spin $i$ and $j$ is given by $J_{ij}$, which will be likewise
parameterized as $J_{ij} = J + \delta J_{ij}$, where $J$ is the
average coupling, and $\delta J_{ij}$ represents random interactions,
being uniformly distributed in $[-\delta J/2,\delta
J/2]$~\cite{remark1}.  Since we are interested in the whole spectrum,
the sign of the exchange coupling parameter $J$ is irrelevant.  In the
numerical simulations, we shall assume $J \geq 0$.  The set of
interacting pairs $\big{\{}i,j\big{\}}$ is determined by the topology
of the lattice.  We shall consider nearest neighbor interactions on a
1D chain and on a 2D rectangular lattice.  In addition, the following
notation is introduced:
\begin{eqnarray}
&&H_Z=\sum_{i=1}^{L} \frac{\varepsilon_i}{2} \sigma_z^{(i)} ,
\label{HZ} \\
&&H_J=\frac{J}{4} \sum_{i=1}^{L-1} \vec{\sigma}^{(i)}
\cdot\vec{\sigma}^{(i+1)}\,.
\label{HJ}
\end{eqnarray}

As discussed in Sec.~IV, these two terms correspond to integrable
limits of the Hamiltonian given in Eq.~(\ref{Ham}).

\subsection{Two-body Random Ensemble}

It is important to appreciate that the class of systems considered
here (Eq.~(\ref{Ham})) is more accurately described by
the so-called {\em two-body random ensemble} (TBRE), which, instead of
$L$-body couplings as implicit in the GOE, involves only two-body
interactions.  The TBRE, and more broadly the embedded Gaussian
ensembles with $k$-body interactions ($k<L$), was introduced in
\cite{French1970,Bohigas1971} as a physically more realistic
statistical setting for describing few-body interacting systems, such
as atoms, molecules, and nuclei.  Among the differences between the
two ensembles \cite{Brody1981,Kota1998,Kota2001}, we highlight the
ones which are most directly relevant to this work:

\begin{itemize}

\item The GOE local density of states (also called the LDOS profile,
see e.g. \cite{Zelevinsky1996,geo1}) as a function of energy shows a
semicircular law, whereas it is Gaussian for the TBRE.

\item The TBRE lacks ergodicity, in the sense that the statistical
properties of each ensemble member need {\em not} coincide with the
ensemble average~\cite{technote}.

\item Contrary to the energy-independent ${\tt NPC}$ value predicted
by the GOE, the ${\tt NPC}$ estimate for the TBRE shows a strong
variation with the energy of the state, approaching the GOE prediction
of $N/3$ mostly in the middle of the energy spectrum.
\end{itemize}

\subsection{Relevant Symmetries}

Although integrability may be regarded as equivalent to the presence
of a {\em complete set of symmetries} of the defining Hamiltonian
\cite{Remark2}, if such a set is incomplete then chaos may still be
present within distinct symmetry sectors. However, a Wigner-Dyson
distribution occurring independently in different sectors will tend to
be washed out if the corresponding energy levels are mixed
together. Thus, in order for clear conclusions to be drawn based on
the energy level statistics, it is necessary to de-symmetrize the
spectrum according to its trivial symmetries, that is, the level
spacing distribution must be examined separately in each resulting
symmetry sector.  The symmetries of the Hamiltonian in Eq.~(\ref{Ham})
are determined by the choice of model parameters $\varepsilon_i$ and
$J_{ij}$:

\begin{enumerate}

\item {\em Rotational symmetry around the $z$-axis.} For all values of
$\varepsilon_i$ and $J_{ij}$, the $z$-component of the total spin,
$S_z = \sum_i \sigma_z^{(i)}/2$, is a conserved quantum number, that
is, $[S_z,H]=0$. For even $L$, the largest subspace corresponds to
$S_z = 0$, having dimension $N_0 ={L \choose L/2}=L!/[(L/2)!]^2$.  We
shall focus primarily on this subspace, ${\cal H}_0$, as chaos sets in
here first.  A natural basis for ${\cal H}_0$ is the one associated
with the eigenstates of $H_Z$, which will be referred to as the {\em
computational basis} or simply c-basis.

\item {\em Conservation of total spin.} When the on-site energies are
degenerate, $\varepsilon_i = \varepsilon$, the total spin, $ S^2 =
(\sum_i \vec{\sigma}^{(i)}/2)^2$, also becomes a good quantum number,
$[S^2,H]=0$. Therefore, in this parameter regime the system must be
separately studied in each symmetry sector characterized by fixed
quantum numbers $S_z$ and $S$~\cite{Remark0}.

\item {\em Symmetries due to lattice geometry.} For $\varepsilon_i =
\varepsilon$ and $J_{ij} = J$, symmetries under site permutations
related to the geometry of the lattice must also be considered.  Under
periodic boundary conditions, $H$ is invariant under cyclic
translations and reflections of the lattice sites, resulting in
momentum and parity conservation respectively.  Note that in
rectangular 2D lattices, momentum and parity in both directions are
good quantum numbers.  For open boundary conditions, the Hamiltonian
is invariant under lattice reflections only.  We shall restrict our
analysis to {\em open boundary conditions} in what follows, because
they lead to fewer symmetries hence larger invariant subspaces, which
permit better eigenvalue statistics. Therefore, when identifying
invariant subspaces, the eigenstates will be grouped according to
their quantum numbers $(S_z,S,R)$, where $R$ indicates parity.

\item {\em Time-reversal invariance.}  GOE is the ensemble appropriate
for describing the static properties of systems exhibiting
time-reversal invariance, that is, for Hamiltonians that commute with
the time-reversal operator, $T_0 = (i)^L \otimes_{j=1}^L\sigma_y^{(j)}
K$, where $K$ is the conjugation operator in the c-basis for
spin-$1/2$. It is important to note that the Heisenberg model with a
magnetic field does {\em not} commute with the conventional time
reversal operator, $T_0$. Nevertheless, the GOE rather that the more
general Gaussian unitary ensemble (GUE) is appropriate for this model
because the Hamiltonian matrix admits a straightforward {\em real}
representation in the c-basis.  Equivalently, one may understand the
applicability of the GOE as resulting from the fact that the
Hamiltonian commutes with $K$ which may be regarded a non-standard
time-reversal operator as in ~\cite{HaakeBook,Avishai2002}.
\end{enumerate}

Although, as mentioned, for a fully established (or weakly broken)
symmetry one may examine the level statistics in each invariant
subspace, for strongly broken symmetry it is no longer possible to
sort eigenvalues by symmetry quantum numbers nor can correlations
between different subspaces be ignored. As a result, separating the
effect of symmetries on the energy level spacing distribution from
that of integrability vs. chaoticity becomes challenging throughout
the symmetry transition regime, that is, it is difficult to
distinguish systematically between the emergence of a complete
vs. incomplete set of symmetries.  While some analytical results exist
for eigenvalue statistics across the full space in the presence of
symmetry-breaking \cite{Guhr2,Hady}, they do not offer a practical
diagnostic tool for distinguishing between fully chaotic and partly
integrable behavior in broken-symmetry regimes based on {\tt LSI}
(especially when multiple symmetry subspaces are involved as in the
$S^2$-symmetry breaking which plays a relevant role here).

\subsection{Related Models}

The Heisenberg model may be regarded as one of a related class of
spin-1/2 lattice models with two-body interaction Hamiltonians of the
form:
\begin{eqnarray}
H_{\tt XYZ} &=& \sum_i \varepsilon_i \sigma_z^i \nonumber \\ &+& \sum
_{\{i,j\}} \left[ J_{ij}^x \sigma_x^{(i)}\sigma_x^{(j)} + J_{ij}^y
\sigma_y^{(i)}\sigma_y^{(j)} + J_{ij}^z \sigma_z^{(i)}\sigma_z^{(j)}
\right]. \nonumber
\end{eqnarray}
In particular, the 2D Ising model in a transverse field $(J^z_{ij} =
J^y_{ij} = 0)$, as well as the XXZ model ($J^x_{ij} = J^y_{ij},
J^z_{ij} \not = 0)$ and the isotropic XY model $(J^x_{ij} =
J^y_{ij},J^z_{ij} = 0)$, have been previously studied in the context
of quantum computation and quantum chaos
\cite{geo2,geo1,lea2,lea3,lea4,Mejia05}.  However, the Heisenberg
model differs in some essential characteristics, notably the
nontrivial role of symmetries.  Also distinctive of the Heisenberg
exchange coupling is the competing nature of the two interacting terms
in the Hamiltonian: in the c-basis, the diagonal Ising interaction
$\sigma_z^{(i)} \sigma_z^{(j)}$ favors localization, whereas the
flip-flop term $\sigma_x^{(i)} \sigma_x^{(j)} + \sigma_y^{(i)}
\sigma_y^{(j)}$ induces delocalization.  Thus, a comparison between
our results and those of previous studies shall also serve to identify
those properties which are generic to disordered spin-$1/2$ lattices
as contrasted to those specific to the Heisenberg model.

\section{Results: Level statistics and delocalization properties}

We shall proceed by first obtaining a basic characterization of the
level spacing distribution and average delocalization properties of
the eigenvectors for a wide range of parameters of interest, and then
proceed to examining entanglement properties in Sec.~VI.

\subsection{From Integrability to Chaos}

The model described by Eq.~(\ref{Ham}) shows two limiting integrable
cases, $H_Z$ and $H_J$ [Eqs.~(\ref{HZ}) and (\ref{HJ}), respectively].
The first is a trivially solvable non-interacting problem, while the
second is solvable only in 1D by Bethe
Ansatz~\cite{Bethe1931,Yang1966,Alcaraz1988,Karbach1998}.  A
transition to chaos may be induced by adding an integrability-breaking
term to any of the two integrable limits. Here, we shall focus on the
following representative scenarios:

\begin{description}
    \item[{\em Case \mbox{\boldmath$ J/\delta \varepsilon$:}}]
    Exchange interactions with a constant strength $J$ are added to
    $H_0=H_Z$.  The crossover between integrability and chaos is
    studied as a function of the ratio $J/\delta \varepsilon$, with
    $\delta J=0$ throughout.

    \item[{\em Case \mbox{\boldmath$\delta J/J $:}}]  Exchange
    interactions with random strength are added to $H_0=H_J$. The
    crossover between integrability and chaos is analyzed as a
    function of the ratio $\delta J/J$, with $\delta \varepsilon=0$
    throughout.

    \item[{\em Case \mbox{\boldmath$\delta J/\delta \varepsilon$:}}]
    Exchange interactions with random strength are added to $H_0=H_Z$.
    The crossover between integrability and chaos is studied as a
    function of the ration $\delta J/\delta \varepsilon$.

\end{description}

Clearly, {\bf {\em Case \mbox{\boldmath$ J/\delta \varepsilon $}}}
includes the other two cases in limiting situations: Specifically, it
coincides with {\bf {\em Case \mbox{\boldmath$ \delta J/\delta
\varepsilon $}}} when the on-site disorder dominates over the disorder
in the coupling strengths, $\delta \varepsilon \rightarrow \infty$,
and with {\bf {\em Case \mbox{\boldmath$\delta J/J$}}} when the
distribution of coupling strengths is very narrow, $J\rightarrow
\infty$.

\subsection{Delocalization Measures}

Because the properties of a state vector (in particular, delocalization)
depend entirely on the choice of representation, quantities such as
${\tt NPC}$ cannot serve as intrinsic (basis-independent) indicators of quantum chaos, such as energy level statistics are considered to be.
This fact may undermine the entire program of establishing connections between the properties of the eigenvectors of a Hamiltonian and its level
of chaoticity.  Yet, it is well-known that for chaotic Hamiltonians, RMT predictions hold (to some extent) in a large number of reasonable representations, although a systematic characterization of such
representations and the degree of agreement one may expect remain at
present an unanswered question.  Here, we conform to the standard
approach to this problem, and examine eigenvector properties with
respect to a basis in which the spread of the eigenvectors has a
physical meaning, for instance one which relates to the relevant
measurement capabilities, or to an integrable limit of the class of Hamiltonians under consideration.  On one hand, we shall investigate
to what extent the eigenvector properties associated with quantum chaos
depend on the choice of two different bases corresponding to the
integrable limits (\ref{HZ})-(\ref{HJ}).  On the other hand,
we shall introduce a measure of relative delocalization between bases
associated with different disorder realizations.  Specifically, we will examine:

\begin{description}

    \item(i) \hspace{0.09cm} The above-mentioned {\em c-basis}, which
is associated with the eigenstates of $H_Z$.  In quantum computation,
this is the basis that represents classical information, relative to
which the final read-out measurement is performed
\cite{Nielsenbook}. The associated ${\tt NPC}$ will be labeled $\xi_c$.

    \item(ii) \hspace{0.03cm} A so-called {\em J-basis}, or
interaction basis, which corresponds to the eigenbasis of $H_J$. In
this case, the associated ${\tt NPC}$ will be denoted by $\xi_J$.

    \item(iii) A disorder-dependent relative representation, which is
obtained as follows. Within a set of sequentially generated random
realizations, the eigenbasis of a given random realization is used to
calculate ${\tt NPC}$ for the eigenvectors of the subsequent
realization.  The associated ${\tt NPC}$, which will be denoted $\xi_r$, quantifies {\em relative delocalization}, that is how
delocalized are the eigenbases associated with different disorder
realizations with respect to each other, rather than with respect to
some fixed basis. By construction, this quantity is independent of any
fixed basis and depends only on the disorder.

\end{description}

Interestingly, an approach which attempts to quantify the
``complexity'' of the eigenstates of an ensemble of Hamiltonians in a
way similar to the above-mentioned relative delocalization has been
proposed based on the notion of \textit{correlational entropy}
\cite{sokolov}.  Both correlational entropy and relative
delocalization overcome the problem of basis dependence by invoking a
distribution over Hamiltonians. An important difference, however, is
that evaluating the correlational entropy requires tracking individual
eigenstates as a function of the disorder across the full disorder
space. While a further comparison between the two concepts would be
worth pursuing, it is likely that this feature could make relative
delocalization computationally more tractable for many disorder
settings.

\subsection{Numerical Results}

We consider the $S_z=0$-subspace, ${\mathcal H}_0$, for 1D and 2D
models. In both cases, for lattice size $L=12$, dim$\,({\cal H}_0)=
N_0=924$.  As anticipated in Sec.~II, ${\tt NPC}$ values are averaged
over all eigenstates in ${\cal H}_0$, and both ${\tt LSI}$ and
${\tt NPC}$ are further averaged over a number of disorder realizations
sufficient for accurate statistics, as indicated in each figure
caption.

\vspace*{1mm}

\begin{figure}[htb]
\includegraphics[width=3.4in]{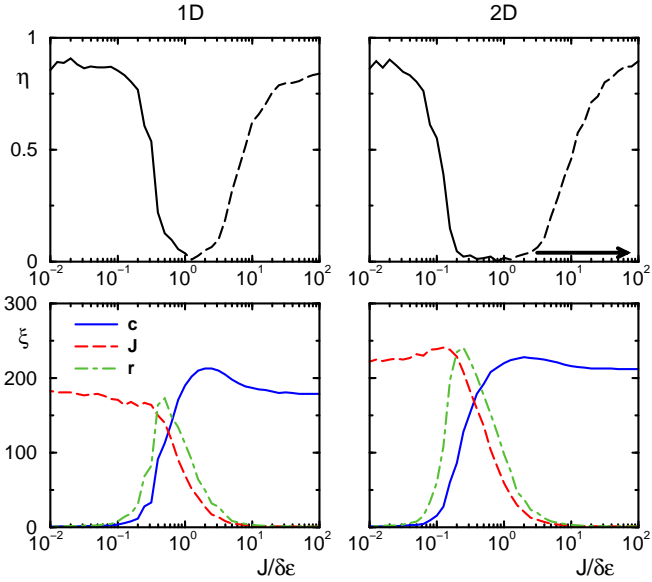}
\caption{ (Color online) {\bf {\em Case \mbox{\boldmath$ J/\delta \varepsilon$}.}}
$S_z=0$ subspace, $L=12$.  Top panels: Level statistics indicator, $\eta$, 
vs. $J/\delta \varepsilon$.  The dashed line indicates in each case 
the transition region, where the $(S,R)$ subspaces are not fully 
distinct.  The arrow indicates the actual limiting value of the 
{\tt LSI} value as $J/\delta \varepsilon \rightarrow \infty$.  
Bottom panels: Number of principal components, $\xi$, 
vs. $J/\delta \varepsilon$.  Left panels: 1D chain, right panels: 2D
$3\times4$ lattice. Averages over 20 random realizations.}
\label{case1}
\end{figure}

{\bf {\em Case \mbox{\boldmath$ J/\delta \varepsilon$}.}} For this
disorder setting, two integrability-chaos transitions are verified for
the 1D model, whereas only one occurs in 2D. From the top panels of
Fig.~\ref{case1}, we see that in both cases the first crossover, from
the $H_Z$ integrable limit to chaos, is observed as the interaction
strength $J$ increases from zero to a value close to the energy
difference between directly coupled states, $J/\delta \varepsilon \sim
1$. In the other extreme, where $J/\delta \varepsilon \rightarrow
\infty $ and therefore $H\rightarrow H_J$, the transition to $S^2$-
and $R$-symmetry complicates the interpretation of the energy level
statistics. During this transition, different $(S,R)$ invariant
subspaces partially overlap, and the observed ${\tt LSI}$ increase
across the whole $S_z = 0$-subspace need not reflect a change towards
integrability, but rather the progressive decoupling of states
belonging to different subspaces (this region is
indicated by a dashed line in the figure).  Once the transition is
complete, at $J/\delta \varepsilon \rightarrow \infty$, the level
spacing distribution within individual $(S_z,S,R)$ symmetry sectors
may be determined. This results in a Poisson distribution for the 1D
model \cite{comment2}, consistent with its exact solvability.
However, and contrary to the behavior of the ${\tt LSI}$ across the
whole ${\cal H}_0$, the level spacing distribution in the invariant
subspaces of the 2D system are strongly Wigner-Dyson.

It is interesting to contrast the ${\tt LSI}$ results with those for
${\tt NPC}$.  The dependence of $\xi_c$, $\xi_J$, and
$\xi_r$ on $J/\delta \varepsilon $ is illustrated in the bottom
panels of Fig.~\ref{case1}.  Corresponding to the broader chaotic
region detected by the ${\tt LSI}$, there is greater delocalization in
2D, especially as quantified by $\xi_J$ and $\xi_r$. For both
cases, the maximum  ${\tt NPC}$ in most instances occurs within the
chaotic region, which we take to be $\eta \lesssim 0.3$
\cite{geo1}.  However, the actual value falls short of the RMT
prediction and different ${\tt NPC}$ measures may disagree
substantially even within the chaotic region.  For example, at
$J/\delta \epsilon \approx 2$ where $\xi_c$ reaches its maximum,
$\xi_J$ and $\xi_r$ show only partial delocalization.
$\xi_c$ and $\xi_J$ detect only one
localized-to-delocalized transition, whereas $\xi_r$ is only
large where ${\tt LSI}$ is low.  Note that the behavior of $\xi_r$
matches more closely the behavior of ${\tt LSI}$ on the low
$J/\delta\epsilon$-$\tt LSI$ transition than $\xi_c$. However,
in the limit $J/\delta\epsilon \rightarrow \infty$ $\xi_r$ fails to discriminate between the integrability of the 1D model and the presence of chaos in the 2D model, decreasing to 1 in both cases due to the absence of disorder.

\vspace*{1mm}

\begin{figure}[t]
\includegraphics[width=3.4in]{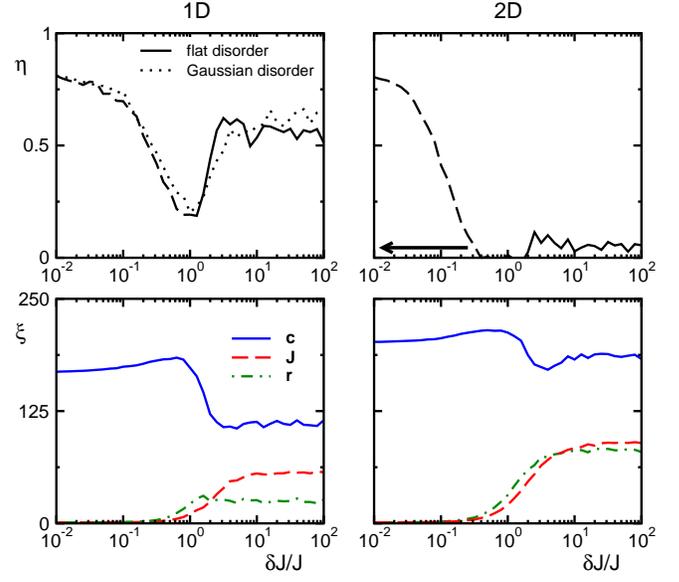}
\caption{(Color online) {\bf {\em Case \mbox{\boldmath$\delta J/J$}}.}
$(S_z,S)=(0,1)$ subspace, $L=12$.  Top panels: Level statistics
indicator, $\eta$, vs. $\delta J/J$.
Dashed lines indicate in each case the region where the
transition to $R$-symmetry sectors occurs. Dotted line: Gaussian
disorder with $\sigma=\delta J/4$. The arrow indicates the actual
limiting value of $\eta$ as $\delta J/J \rightarrow 0$. Bottom
panels: Number of principal components, $\xi$, vs. $\delta J/J$.
Left panels: 1D chain, right panels: $3\times4$ lattice.
Averages over 50 random realizations.}
\label{case3}
\end{figure}

{\bf {\em Case \mbox{\boldmath$\delta J/J$}}.} In this case,
$[S^2,H]=0$, thus we examine ${\tt LSI}$ and ${\tt NPC}$ in the largest
$S^2$-subspace, corresponding to $S=1$ and dimension $N=297$
eigenstates. As $\delta J/J \rightarrow 0$ ($H\rightarrow H_J$), the
reflection symmetry becomes important, causing the ${\tt LSI}$ in the
top panels of Fig.~\ref{case3} to increase irrespective of the
chaoticity or integrability of the Hamiltonian. As previously
discussed, in this limit the 1D model is integrable, while the 2D
model is not. Consequently, as $\delta J/J$ increases from 0, a
transition from integrability to chaos occurs for the 1D model,
whereas only the breaking of the reflection symmetries occurs in the
2D model. Notably, for the 1D model near $\delta J /J =2$, an {\em
unexpected rapid transition} in ${\tt LSI}$ from $\eta \sim 0.2$ to an
intermediate value of $\eta \sim 0.6$ is observed.  Although the 2D system
remains chaotic throughout, a small rise in ${\tt LSI}$ close to
$\delta J / J \sim 2$ is noticeable.

{\tt NPC} values are shown in the bottom panels of
Fig.~\ref{case3}.  Since the Hamiltonians in the J-basis and in
any basis associated to a disorder realization are block-diagonal,
${\xi_J}$ and ${\xi_r}$ cannot exceed the dimension of the
$(S_z,S)=(0,1)$ subspace, unlike ${\xi_c}$ which is upper-bounded
only by the dimension of the $S_z=0$ subspace given that states of the
c-basis do not posses $S^2$-symmetry.  For the 1D model, the slight
increase in ${\xi_c}$ as $\delta J/J \rightarrow 1$ occurs during
the integrability-chaos transition, but the same occurs also for the
2D system, which is chaotic throughout.  Thus, this increase is likely
related to the reflection symmetry-breaking. Interestingly, the abrupt
drop in ${\xi_c}$ for both 1D and 2D near $\delta J/J = 2$ appears
to be connected with the rise in ${\tt LSI}$ -- the greater change in
both ${\tt LSI}$ and ${\xi_c}$ occurring for the 1D model.

A possible explanation for the observed transition at $\delta J/J\sim
2$ is based on the fact that this value marks the point where
interaction strengths arbitrarily close to zero are allowed, $\min_{ij} \{ J_{ij}\} =J-\delta J/2=0$. If the chain is broken into two or
more approximately uncoupled segments, then the energy spectrum of
each segment becomes approximately independent.  Even if for each
segment the level spacing distribution is Wigner-Dyson, the level
spacing distribution for the combined spectrum will not be. Lending
support to this explanation is the far stronger effect in 1D than in
2D, following from the fact that the 2D lattice cannot be as easily
broken into isolated segments.  In order to determine if this effect is an artifact of the sharp cut-off in the disorder distribution, interactions with Gaussian disorder (with standard deviation $\sigma=\delta J/4$) were also examined. The results are shown as the dotted line in Fig.~\ref{case3} (top left), where it is seen that the abrupt transition remains despite the absence of a sharp cut-off in the disorder range (note, however, that the width of the transition is affected).

It is intriguing to notice that in 1D the ground state and low-energy
excitations for this disorder setting have been determined at $\delta
J/J = 2$ through an exact renormalization group approach, and are
known to be described by the so-called {\em random singlet
phase}~\cite{Ma,Dasgupta,Bhatt}. A more thoroughly discussion of
possible connections between the observed behavior and the existence
of a random singlet phase are left for future investigation.

As a further remark, the behavior of $\xi_r$ in the
region where $\delta J/J>2$ is worth mentioning.  In 1D, $\xi_r$
is much lower than $\xi_J$, while in 2D only a small difference is
seen. This further demonstrates that, in the presence of disorder, {\em
relative delocalization can be an effective indicator of chaos}.

\vspace*{1mm}

\begin{figure}[t]
\includegraphics[width=3.4in]{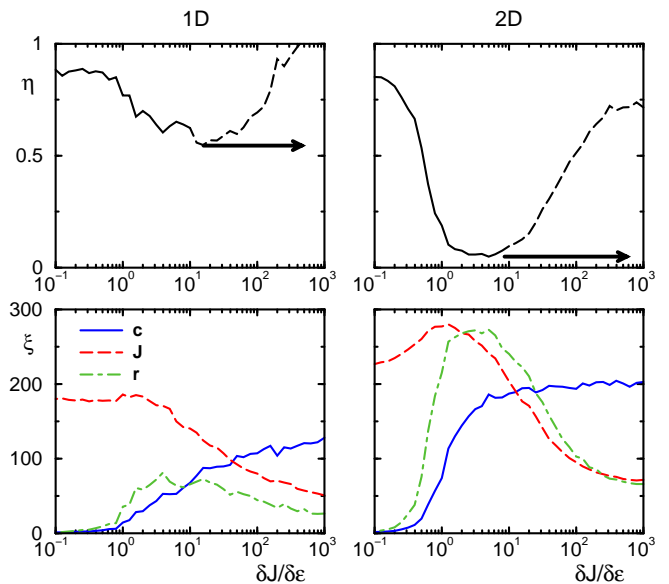}
\caption{(Color online) {\bf {\em Case \mbox{\boldmath$\delta J/\delta
\varepsilon$}}.}  $S_z=0$ subspace, $L=12$.  Top panels:
Level statistics indicator, $\eta$, vs. $\delta J/\delta \varepsilon$.
Dashed line indicates the region where the transition to
$S^2$-symmetry sectors occurs.  Arrow indicates the actual
limiting value of $\eta$ as $\delta J/\delta
\varepsilon \rightarrow \infty$.  Bottom panels:
Number of principal components, $\xi$, vs. $\delta
J/\delta \varepsilon$.  Left panels: 1D chain, right panels:
$3\times4$ lattice. Averages over 50 random realizations.}
\label{case2}
\end{figure}

{\bf {\em Case \mbox{\boldmath$\delta J/\delta \varepsilon$}}.}  The
dependence of ${\tt LSI}$ on $\delta J/\delta \varepsilon$ is shown in
the top panels of Fig.~\ref{case2}. As the disorder in the coupling
strengths becomes sufficiently large to compete with the energy
difference $\delta \varepsilon$, a transition from integrability to
fully developed chaos is observed only for the 2D model, whereas in 1D
the${\tt LSI}$ merely approaches an intermediate value of $\eta \sim
0.6$~\cite{comment}. For $\delta J/\delta \varepsilon \gtrsim 10$,
${\tt LSI}$ for the entire $S_z=0$ subspace increases, but, as before,
this does not necessarily reflect an approach to integrability, since
a transition to the $S^2$-symmetry is in place. In fact, at $\delta
J/\delta \varepsilon \rightarrow \infty$, we have verified that
essentially the same ${\tt LSI}$ values reached at $\delta J/\delta
\varepsilon \sim 10$ for 1D and 2D within the $(S_z,S)=(0,1)$ subspace
are attained.

The ${\tt NPC}$ behavior is depicted in the bottom panels of
Fig.~\ref{case2}.  Reflecting the fact that chaos never fully develops
in 1D, both ${\xi_c}$ and ${\xi_r}$ are considerably lower
when compared to the previous case.  In contrast, for the 2D model,
this disorder setting leads to a high level of delocalization,
especially in the J-basis and in terms of relative delocalization.
The maximum values are reached between $1\lesssim \delta J/\delta\varepsilon \lesssim 6$. As $\delta J/\delta \varepsilon$ increases further, the
transition to $S^2$-symmetry sectors becomes relevant, and the
Hamiltonians in these two bases approach a block-diagonal form --
therefore ${\xi_J}$ and ${\xi_r}$ decrease accordingly.  As
$\delta J/\delta \varepsilon \rightarrow \infty $, the values obtained
still indicate strong delocalization relative to the sizes of the
$(0,S)$ subspaces.

\begin{figure}[t]
\includegraphics[width=2.5in]{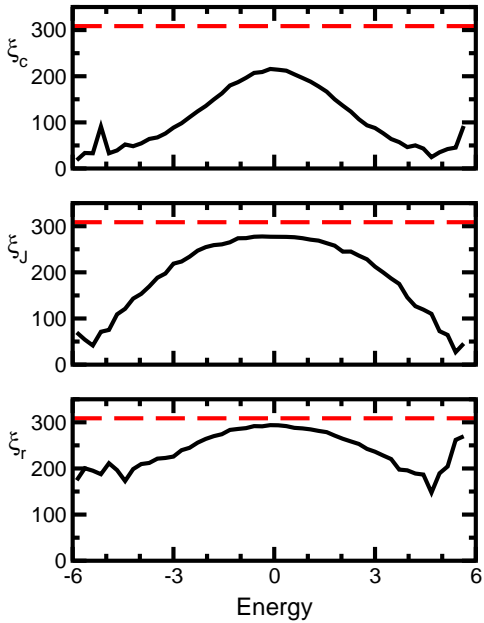}
\caption{ (Color online) {\bf {\em Case \mbox{\boldmath$\delta J/\delta
\varepsilon$}}.}  Number of principal components vs. energy 
eigenvalue for $\delta J/\delta \varepsilon=3$. Eigenstates in the
$S_z=0$-subspace for a $3\times4$ qubit lattice are considered.
The dashed horizontal line shows the
GOE prediction, $\xi_{\rm GOE}=(N_0+2)/3 \sim 308$.  Averages
over 20 random realizations.}
\label{NPCvsE}
\end{figure}

Consistent with studies of the TBRE and of other systems with two-body
interactions (notably, nuclear-shell models), a strong dependence of
${\tt NPC}$ on energy is observed, particulary in chaotic regions, see
Fig.~\ref{NPCvsE}.  Specifically, the ${\tt NPC}$ follows an
approximately Gaussian form. We show a representative 2D case at
$\delta J/\delta \varepsilon \sim 3$, where ${\xi_J}$ and ${\xi_r}$ are close to their maximum. Note that RMT provides a bound on
the extent of delocalization, and that even in the center of the
spectrum, the average delocalization falls short of the predictions of
RMT, particularly in the c-basis. That the eigenstates in the tails of
the spectrum are less delocalized than those in the center, however,
only partially affects the average ${\tt NPC}$'s shown in
Figs.~\ref{case1}, \ref{case3}, \ref{case2}, because the density
of states is also peaked around the center of the spectrum.  We found
that averaging only over the central 100 eigenstates ($10.6 \%$ of the
total) increases the values of each ${\tt NPC}$~-measure by an amount between
10\% to 15\%, and does {\em not} qualitatively affect the behavior of
the average ${\tt NPC}$. For all cases except for ${\xi_J}$ and
${\xi_r}$ in case $\delta J/\delta\epsilon$ near their maxima, the
deviation from the RMT prediction of ${N}/{3}$ remains substantial.

This situation is to be compared with the models studied in
\cite{geo1} and \cite{Mejia05}.  Even though these models also correspond to the TBRE as well, better agreement with RMT is obtained for
delocalization whenever the interactions are either purely {\it off-diagonal} or {\em all-to-all}, that is, every site is coupled with every other site.
In fact, a key difference in our case is the presence of the {\it purely diagonal} Ising contribution, which favors localization in the computational basis.  Furthermore, we have verified that all-to-all Heisenberg couplings also yield delocalization values in better agreement with RMT (data not shown).  Thus, the lower level of ergodicity in the models examined
here may be partly attributed to lower connectivity.

\section{Entanglement Measures}

Obtaining a complete characterization of entanglement in many-body
systems is a challenging problem which in spite of extensive effort
remains as yet largely unsolved \cite{Fazio07}.  As a main reason for
such difficulty, it is evident that no single entanglement measure can
fully capture the complexity of multi-particle correlations.  For our
current purposes, we shall select representative measures of
entanglement that are both computationally practical, and directly
connect with prior work on entanglement and quantum chaos and/or
quantum phase transitions. Specifically, we shall investigate
so-called concurrence between selected pairs of qubits as a measure of
pairwise correlations~\cite{Woot}, and a family of multipartite purity
measures constructed within the general GE
framework~\cite{BKOV03,BKOSV04}.

For a pure state $\ket{\psi}$ of two qubits, {\em concurrence} may be
defined as
$$C(|\psi\rangle)=|\bra{\psi}\sigma_y^{(1)} \otimes \sigma_y^{(2)}
\ket{\psi^*}|,$$
\noindent
where $\ket{\psi^*}$ is the complex conjugate of $\ket{\psi}$.
Physically, $C(|\psi\rangle)$ may be thought of in this case as the
overlap of a state with its time-reversed counterpart.  When a pure
state $|\psi\rangle$ of $L>2$ qubits is considered, the reduced
two-qubit state of a selected pair ($i,j)$ is described by a density
matrix $\rho_{ij}$, where all but the qubits of interest are traced
out.  In this case, concurrence may be computed through a more general
expression, which holds also for {\em mixed} states and reads
\begin{equation}
C(\rho_{ij}) = \max(0,\lambda_1- \lambda_2 -
\lambda_3 - \lambda_4),
\end{equation}
where $\lambda_1 \geqslant \lambda_2 \geqslant \lambda_3 \geqslant
\lambda_4$ are the square roots of the eigenvalues of $\rho_{ij}
\tilde{\rho}_{ij}$, and $\tilde{\rho}_{ij} =
(\sigma_y^{(1)}\otimes\sigma_{y}^{(2)})$ $\rho_{ij}^*
(\sigma_y^{(1)}\otimes\sigma_y^{(2)})$ \cite{Woot}.  Concurrence
ranges from a minimum of $0$ for unentangled states to a maximum of
$1$ for states containing a maximum amount of pairwise correlations.
Thus, applied to a pair of qubits in a pure many-qubit state, zero
concurrence will occur for both a product state, $\ket{\psi} =
\ket{\phi_1} \otimes ... \otimes \ket{\phi_L}$, but also for the
$L$-partite Greenberger-Horne-Zeilinger state, $\ket{{\tt GHZ}_L} =
\frac{1}{\sqrt{2}}(\ket{0,...,0} + \ket{1,...,1})$, which exhibits
genuinely multipartite correlations~\cite{GHZ}. For random states of a
sufficiently large number of qubits, the expected value of concurrence
between any two qubits is zero~\cite{Scott}.

For a multipartite system, concurrence quantifies the amount of
mixed-state bipartite entanglement within a given pair.  Complementary
information about how the pair itself or, more generally, a given
subset $A$ of qubits, correlates with the remaining subset $B$, is
provided by the amount of bipartite entanglement between $A$ and $B$.
The unique measure satisfying all requirements of invariance under
local transformations, continuity, and additivity is the {\em von
Neumann entropy} of either reduced density matrix, e.g.,
\[ S (|\psi\rangle_{AB})= - {\rm Tr}_A (\rho_A \log_2 \rho_A), \]
where $\rho_A$ is the reduced density operator of subsystem $A$.  If
the additivity requirement is relaxed, a simpler linearized version
of the above expression suffices to quantify bipartite entanglement,
the so-called {\em linear entropy},
\[ E (|\psi\rangle_{AB})= 1 - {\rm Tr}_A (\rho_A)^2. \]
Although the amount of multi-partite entanglement is generally hard to
quantify away from bipartite states, useful indirect insight may be
gained by considering different bi-partitions of the system.  In
particular, $|\psi\rangle$ certainly contains genuine multipartite
entanglement if no reduced subsystem state is pure.

The notion of GE offers a powerful framework for both organizing
conventional multipartite entanglement within a unified setting, and
for extending the concept of entanglement to situations where a
preferred partition of the system into subsystems may not be
meaningful or otherwise desirable
~\cite{BKOV03,BKOSV04,SOBKV04,allthat}.  GE is based on the
relationship of the state of interest to a {\em distinguished set of
observables}, rather than to a tensor product decomposition of the
Hilbert space ${\cal H}$ into subsystems.  Let such a distinguished
observable set consists of the Hermitian operators in a linear
subspace $h$ of the full operator space on ${\cal H}$, with $h$ closed
under Hermitian conjugation.  The key step is to replace the notion of
subsystem state as obtained via the usual partial trace operation by a
notion of ``reduced state'' as determined by the expectation values of
observables belonging to the restricted subspace $h$.  A pure state
$|\psi\rangle$ is {\em generalized unentangled (entangled) relative to
$h$} depending on whether its reduced state is pure (mixed) in the
space of all reduced states that is, depending on whether it is {\em
extremal} (or not) in the convex sense.  If $\{B_i\}$ is a basis of
Hermitian traceless operators for $h$, orthogonal in the trace inner
product, a natural measure for GE is the degree of purity of the
reduced $h$-state as quantified by the so-called {\em $h$-purity},
\begin{equation}
P_h(|\psi\rangle)= \kappa \sum_i |\bra{\psi}B_i\ket{\psi}|^2,
\label{hpur}
\end{equation}
where the overall normalization constant $\kappa$ is chosen so that
$P_h$ ranges between its maximum value of $1$ for generalized
unentangled states and its minimum value of $0$ for maximal GE
relative to $h$.

When the observable set $h$ is the Lie algebra of all local
observables on qubits, GE reduces to standard multipartite qubit
entanglement.  In particular, the above $h$-purity (\ref{hpur})
relative to arbitrary single-qubit observables takes the explicit
form:
\begin{equation}
P_1(\ket{\psi})= \frac{1}{L}
\sum_{\alpha=x,y,z}^{i=1,L}
|\bra{\psi}\sigma_\alpha^{(i)}\ket{\psi}|^2.
\end{equation}
This quantity is simply related to the so-called Meyer-Wallach measure
of global entanglement, $Q$, by $P_1=
1-Q$~\cite{MeyerWallach,SOBKV04,Brennen}, which quantifies
multipartite entanglement through the {\em average} bipartite
entanglement between a qubit and the rest.  By construction, $P_1=1$
for product states, whereas $P_1=0$ for states such as $\ket{{\tt
GHZ}_L}$, where each single-qubit reduced density matrix is totally
mixed.

\begin{figure}[b]
\includegraphics[width=3.in]{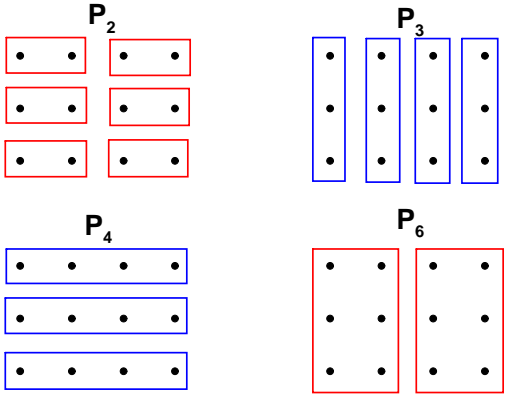}
\caption{(Color online) Choices of partitions for evaluating 
$n$-local purities in the $3\times4$ lattice.}
\label{fig:2D}
\end{figure}

A family of related entanglement measures may be naturally constructed
from the above local purity by coarse-graining.  If the qubits are
partitioned into distinct $n$-qubit blocks, the {\em $n$-local
purity}, denoted $P_n$, is defined as purity with respect to arbitrary
observables local to each block.  For example, the bi-local purity
(also used in~\cite{Montangero2006}) is given by
\begin{equation}
P_2(\ket{\psi})= \frac{2}{3L}
\sum_{\alpha,\beta=x,y,z}^{i=1,L/2}
|\bra{\psi}\sigma_\alpha^{(2i-1)}\sigma_\beta^{(2i)}\ket{\psi}|^2,
\end{equation}
where the sum only extends to traceless operators.  Physically, the
measure $P_n$, for $n>1$, ignores short-range correlations which $P_1$
detects. For the 1D chain we consider, the natural choice of
partitions is into contiguous blocks. For the 2D lattice, relevant
block partitions are illustrated in Fig.~\ref{fig:2D}.

It is important to mention that the von Neumann entropy of an
$n$-qubit block, which has been a more broadly used quantity in
studies of entanglement vs quantum
chaos~\cite{Bandyopadhyay2002,Bandyopadhyay2004,Mejia05,Lakshminarayan2006},
is closely related in meaning to $P_n$. It can be shown that the
average bipartite entanglement between the $n$-qubit block and the
rest, $Q_n=1-P_n$, is (up to a normalization constant) equal to the
average linear entropy over each $n$-qubit block participating in
$P_n$.  Thanks to the average over many blocks, $P_n$ is less
sensitive to edge effects.  Apart from that, a main advantage with
respect to von Neumann block entropy is its mathematical simplicity,
which allows analytic calculations of the expected GE for random pure
states to be and also reveals a quantitative relationship with
delocalization as measured by ${\tt NPC}$ in a local basis~\cite{pvn}.
In particular, for random pure states with purely {\em real}
components, as expected for GOE eigenvector statistics, the average
value $\overline{P_n}$ with respect to the appropriate Haar measure
may be exactly computed for arbitrary lattice size $L$.  While we
defer the details of the derivation to Appendix A, the results for
states in the $S_z=0$-subspace with $L=12$ are summarized in Table I.

\begin{table}[tb]
\begin{center}
\begin{tabular}{ccccc}
$\hspace{0.3cm} \overline{P_1} \hspace{0.3cm}$ &
$\hspace{0.3cm} \overline{P_2} \hspace{0.3cm}$ &
$\hspace{0.3cm} \overline{P_3} \hspace{0.3cm}$ &
$\hspace{0.3cm} \overline{P_4} \hspace{0.3cm}$ &
$\hspace{0.3cm} \overline{P_6} \hspace{0.3cm}$ \\\hline
$2.16$ & $5.69$ & $7.71$ &
$7.49$ & $10.10$
\end{tabular}
\end{center}
\caption{Expected $n$-local purity for random states in the $S_z=0$
subspace of a $L=12$-site lattice, $N_0=924$.  Values are multiplied
by a factor $\times10^{3}$ for clarity.}
\end{table}

The intrinsic flexibility of the GE notion easily allows for the
construction of purity measures relative to distinguished operator
spaces not directly tied to a subsystem partition.  For spin-$1/2$
systems, in particular, a natural choice of distinguished observables
emerges upon mapping Pauli spin operators to canonical fermionic
operators via the Jordan-Wigner transform,
$$c_j = \Big( \otimes_{i=1}^{j-1} \sigma_z^{(i)} \Big)
\sigma_+^{(j)},\;\;\;\; \sigma_+^{(j)}=\frac{1}{2}(\sigma_x^{(j)} + i
\sigma_y^{(j)}).$$
\noindent
It is straightforward to show that the $\{c_i\}$ satisfy canonical
fermionic anti-commutation relations, $\{c_i^\dagger , c_j\} =
\delta_{ij}$, $\{c_i, c_j\} = 0$.  Then ``generalized local''
resources may be associated with quadratic fermionic observables
commuting with the total fermionic number operator, $\hat{L}=\sum_i
c^\dagger_i c_i$ -- as opposed to ``nonlocal'' resources involved in
processes where the total fermion number in a given pure state may
change.  The corresponding distinguished observable set is isomorphic
to the unitary Lie algebra in $L$ dimensions, $h=\mathfrak{u}(L)$.  GE
relative to such algebra is quantified by the fermionic
$\mathfrak{u}(L)$-purity \cite{SOBKV04},
\begin{eqnarray}
P_{\mathfrak{u}(L)} (|\psi\rangle) &=& \frac{2}{L} \sum_{i <j=1}^L
\Big ( \langle c^\dagger_i c_j +c^\dagger_j c_i\rangle^2 - \langle
c^\dagger_i c_j -c^\dagger_j c_i\rangle^2 \Big) \nonumber \\ & +&
\frac{4}{L}\sum_{i =1}^L \langle c^\dagger_i c_i -\frac{1}{2}\rangle
^2 \,.
\label{fermP}
\end{eqnarray}
Physically, $P_{\mathfrak{u}(L)}$ indicates how ``close" a state is to
being described by a fermionic product state (a Slater determinant)
\cite{allthat}.  For example, in the two excitation sector,
$P_{\mathfrak{u}(L)}(\ket{\psi})=1$ if and only if $\ket{\psi}$ may be
written in the form $\ket{\psi} = c_a^\dagger c^\dagger_b |{\tt
vac}\rangle$, where $a,b$ label any set of modes unitarily related to
modes $i,j$, and $|{\tt vac}\rangle$ contains no
fermions. $P_{\mathfrak{u}(L)}$ has been shown to successfully detect
and characterize broken-symmetry quantum phase transitions
\cite{SOBKV04}.  In the context of the transition to quantum chaos,
such a measure may be expected to provide insight into entanglement
generation associated with the departure from a non-interacting
fermion problem.

\section{Results: Entanglement behavior}

We are now in a position to address entanglement properties of
Heisenberg-model eigenvectors, and to understand the resulting
behavior based on the insight gained from the corresponding behavior
of delocalization and energy level statistics.

\subsection{Disorder Dependence}

{\bf {\em Case \mbox{\boldmath$ J/\delta \varepsilon$}.}}  As
remarked, this setting is especially useful for exploring possible
connections between integrability-breaking, delocalization, and
entanglement, because it contains both the integrable limits
associated with $H_Z$ (in 1D and 2D) and $H_J$ (in 1D).  With respect
to the c-basis, the 1D model possesses three regimes of interest: (i)
integrability-localization for $J/\delta \varepsilon \rightarrow 0$;
(ii) chaoticity for intermediate ratios $J/\delta \varepsilon \sim 1$;
and (iii) integrability-{\em de}localization for $J/\delta \varepsilon
\rightarrow \infty$ in 1D, while in latter regime the 2D model remains strongly non-integrable.

\begin{figure}[b]
\includegraphics[width=3.4in]{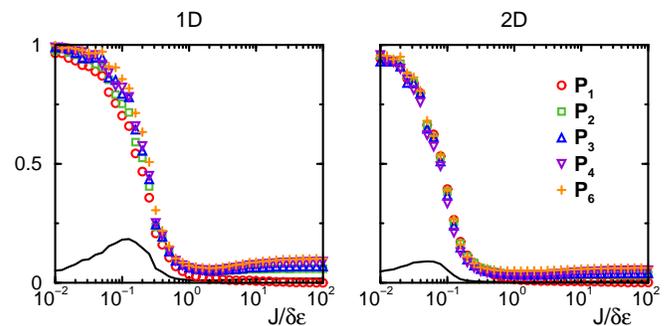}
\caption{(Color online) {\bf Case} \mbox{\boldmath$ J/\delta \varepsilon$}. 
Average nearest-neighbor concurrence and $n$-local purity averaged over all
eigenvectors of the $S_z=0$-subspace vs $\delta J/\delta
\varepsilon$. Lattice size $L=12$. Solid black line: concurrence.
Left panel: 1D chain, right panel: $3\times4$ lattice.  Averages over
20 random realizations.}
\label{m2d}
\end{figure}

Fig.~\ref{m2d} summarizes the behavior of the average concurrence
between all neighboring spins, $C$, and of the $n$-local purities,
$P_n$, for the 1D and 2D models. As we depart from the purely local
Hamiltonian $H_Z$ and $\delta J/\delta \varepsilon$ increases from 0
to 1, a peak in $C$ is observed {\em before} the onset of chaos. This
suggests that, as the system delocalizes, pairwise entanglement is
the first type of entanglement to emerge -- disappearing, however, in
the chaotic region \cite{lea3}.  Such a disconnection between pairwise
entanglement and the onset of chaos has also been verified for the 2D
Ising model in a transverse field in \cite{Mejia05} and, more
recently, \cite{Lakshminarayan2006}.  Contrary to that, each of the $P_n$ decreases from nearly 1 to nearly 0, roughly corresponding to the transition in {\tt LSI} and $\xi_c$ (cf. Fig.~\ref{case1}).  Therefore, as we approach chaoticity, a shift from pairwise to genuinely multipartite entanglement occurs. Interestingly, in 1D a shift from short-range to long-range correlations is also noticeable, since a more pronounced
decay is witnessed by $P_1$, which is subsequently followed by the
other purities until finally being reached by $P_6$.  In 2D, different
$P_n$ do not directly signify correlations over different distance,
and all $P_n$ curves superimpose more closely.

\begin{figure}[b]
\includegraphics[width=3.4in]{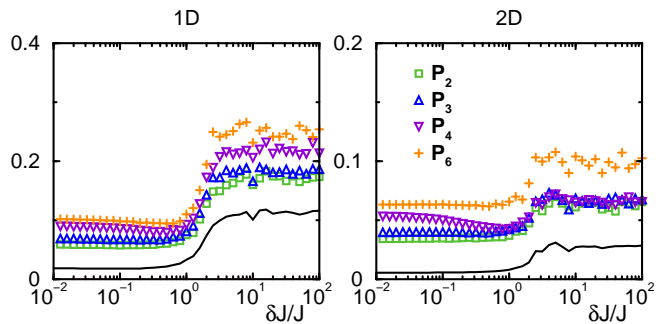}
\caption{(Color online) {\bf {\em Case \mbox{\boldmath$\delta J/J$}}.}  Average
nearest-neighbor concurrence and $n$-local purity averaged over all
eigenvectors of the $(S_z,S)=(0,1)$ subspace vs $\delta J/J$.  Lattice
size $L=12$.  Solid black line: concurrence.  Left panel: 1D chain,
right panel: $3\times4$ lattice.  Averages over 20 random
realizations.}
\label{ms}
\end{figure}

In the region where $J/\delta \varepsilon$ increases from 1 to
$\infty$ and the limiting Hamiltonian $H_J$ is approached, $C$ and
$P_1$ are near zero throughout -- note that due to the rotational
invariance of $H_J$, $P_1 \rightarrow 0$ as $J/\delta \varepsilon
\rightarrow \infty$, whereas $P_2, P_3, P_4$, and $P_6$ increase
slightly, particularly in 1D. It is hard to ascertain whether this
subtle variation in 1D is related to a transition to integrability or
it is simply associated to the symmetry transition.  Therefore, the
similar entanglement behavior observed in both 1D and 2D does not
mirror the differences in chaoticity between the two systems.  In fact,
it appears much more closely connected to the $\xi_c$ behavior
in this region.  In this sense, a parallel between standard
multipartite entanglement and delocalization is stronger than between
the former and chaos.  Furthermore, similar to delocalization, the minimum values reached by each $P_n$ in the central chaotic region, although low, do not attain the values predicted by RMT. For the 2D case, where agreement with RMT is best, the minimum of each $P_n$ except for $P_1$ range between a factor of 4.5 to 5.8 times the RMT values in Table I.
$P_1$ does attain the RMT value but only well into the $S^2$ symmetry region, where this effect cannot be attributed to chaos.

{\bf {\em Case \mbox{\boldmath$\delta J/J$}}.} Similar to the above
Case $J/\delta \varepsilon$ in the region $J/\delta \varepsilon \geq
1$, the entanglement behavior shows no qualitative difference between
the 1D and 2D models for $\delta J/J \in [0,1]$, although an
integrability-chaos transition occurs in 1D (see Fig.~\ref{ms}).  On
the other hand, both $P_n$ and $C$ change abruptly near $\delta J/J
\sim 2$ in 1D, and near $\delta J/J \sim 3$ in 2D.  In both cases, the
sharp increase in pairwise entanglement and decrease in multipartite
entanglement are paralleled by a decrease in $\xi_c$ and, more
interestingly, by an increase in $\tt LSI$ (cf.
Fig.~\ref{case3}). It is expected that the breaking of the chain
into distinct subsystems will result in a decrease of multipartite
entanglement.  It is, however, intriguing that although smaller, a
decrease occurs also in 2D, where chain breaking should be suppressed.

{\bf {\em Case \mbox{\boldmath$\delta J/\delta \varepsilon$}}.}  The
entanglement behavior in 2D for this disorder setting is very similar
to the one depicted in the right panel of Fig.~\ref{m2d}, whereas in
1D purities never reach low values, reflecting the lack of
delocalization (cf.  Fig.~\ref{case2}).  We shall then focus on the 2D
system and describe some interesting features of purity, which also
apply to Case $J/\delta \varepsilon$ in both 1D and 2D.

\subsection{Energy and Delocalization Dependence}

In addition to understanding how entanglement properties depend on
disorder and interaction strength, their direct dependence upon energy
and delocalization is relevant to gaining a more complete physical
picture.

\begin{figure}[t]
\includegraphics[width=3.4in]{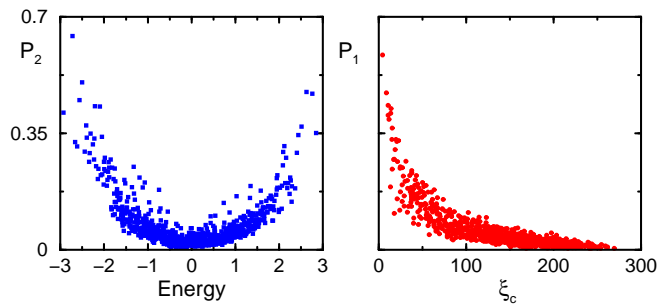}
\caption{ (Color online) {\bf {\em Case \mbox{\boldmath$\delta J/\delta
\varepsilon$}}.} $S_z=0$ subspace, 3x4 2D spin lattice, $\delta J/
\delta \varepsilon=2$.  Left panel: $P_2$ vs. energy eigenvalue.
Right panel: $P_1$ vs $\xi_c$.  A single disorder realization is 
considered. }
\label{p2e}
\end{figure}

Similarly to the behavior observed for ${\tt NPC}$ within the TBRE
(see Fig.~\ref{NPCvsE}), a strong correlation also exists between
multipartite entanglement and the energy spectrum. The left panel of
Fig.~\ref{p2e} contains a representative example for the bi-local
purity.  In general, states at the edges of the spectrum tend to be
less entangled, whereas highly entangled states are clustered around
intermediate energies.  This result indicates a correspondence between
$P_n$ and the LDOS of the TBRE, which, as a function of energy, is
Gaussian and broadly peaked at the center of the spectrum
\cite{Brody1981,Kota1998,Kota2001} (see also \cite{Montangero2006} for
a discussion of the influence of LDOS properties on purity behavior).
Essentially, where the LDOS is largest, a higher level of
delocalization exists and also higher amounts of multipartite
entanglement occur.  We emphasize, however, that this relationship is
not {\em per se} indicative of chaos: A similar behavior was found for
disordered on-site energies (Case $J/\delta \varepsilon$) within the
$(S_z,S,R)=(0,1,+1)$ sector for the 1D model in the limit $J/\delta
\varepsilon \rightarrow \infty$, which corresponds to a delocalized,
but {\em integrable} regime.

The relationship between multipartite entanglement and delocalization
may further be probed by directly comparing $n$-local purities and
${\tt NPC}$.  A plot of the local purity, $P_1$, as a function of
delocalization in the c-basis (right panel of Fig.~\ref{p2e})
discloses a striking relationship between $P_1$ and $\xi_c$,
which is found to persist for each one the $P_n$ and for a broad range of
values of $\delta J/\delta \varepsilon$ within the chaotic region
(until the $S^2$-symmetry becomes strong).  Remarkably, for a given
$P_n$, the precise shape of the curve does {\em not} depend on $\delta
J/\delta \varepsilon$, although it depends slightly on lattice
dimension and disorder setting.  The precise relationship between
$P_1$ and $\xi_c$ is examined at length in \cite{pvn}, where it
is shown that the relationship between $P_1$ and $\xi_c$ depends
strongly on how correlated the state vector components are with
respect to the {\em Hamming distance} between the quantum numbers
describing c-basis states~\cite{Hamming}.  If no such
Hamming-correlation exists, inverse proportionality between local
purity and ${\tt NPC}$ is expected (see also \cite{Montangero2006}):
$$ P_1 (|\psi\rangle) = \frac{N}{N-1}\frac{1}{\xi_c
(|\psi\rangle)} - \frac{1}{N-1}\approx \frac{1}{\xi_c
(|\psi\rangle)},$$
\noindent
with $N=N_0=924$ for the central band.  While this qualitatively
agrees with the observed behavior, due to the fact that all
interaction terms in the Hamiltonian are of a two-body nature, the
components of the eigenvectors tend to be Hamming-correlated,
resulting in significant deviations from the predicted inverse scaling
law, especially at small $\xi_c$.  This effect was recently
independently confirmed by Giraud and coworkers \cite{Giraud07}.

\begin{figure}[t]
\includegraphics[width=3.4in]{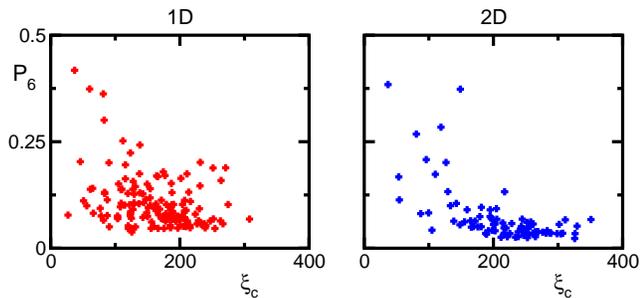}
\caption{(Color online) Multipartite entanglement vs. delocalization in a 
clean Heisenberg Hamiltonian: $P_6$ vs $\xi_c$ in the
$(S_z,S,R)=(0,1,+1)$ symmetry subspace.  Lattice size: $L=12$.  Left
panel: 1D chain. Right panel: $3\times 4$ 2D lattice.}
\label{P6vsNPC}
\end{figure}

Beside calling for a deeper understanding of the physical conditions
leading to the observed non-trivial eigenvector structure, the above
findings naturally prompt the following question: To what extent could
the relationship between entanglement and delocalization provide a
signature of quantum chaos?  As a first step toward answering this
question, we revisit the case of a {\em clean} Heisenberg Hamiltonian,
$H_J$ (Case $J/\delta \varepsilon$ in the limit of no disorder), which
supports both integrability (in 1D) and chaoticity (in 2D).  Instead
of $P_1$, which is identically zero for eigenvectors of $H=H_J$, we
look at the block-purity $P_6$.  As illustrated in Fig.~\ref{P6vsNPC},
no clear relationship between $\xi_c$ and $P_n$ emerges in the
integrable regime (left panel), whereas a noticeable relationship is
present in the chaotic case (right panel) -- in spite of the limited
statistics accessible due to symmetry constraints.  Interestingly,
this is in agreement with a recent study of entanglement and chaos for
the Ising model in a transverse field by Lakshminarayan and coworkers
\cite{Lakshminarayan2006}, where a correlation between values of
delocalization and block von Neumann entropy is found to occur only in
chaotic regimes and not in integrable yet delocalized ones.

\begin{figure}[t]
\includegraphics[width=3.4in]{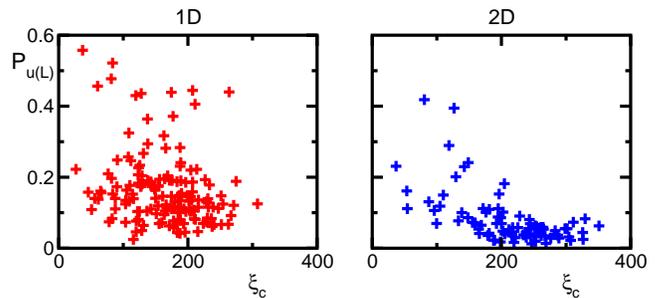}
\caption{(Color online) Generalized fermionic entanglement vs delocalization 
in a clean Heisenberg Hamiltonian: $P_{u(L)}$ vs. $\xi_c$ in the
$(S_z,S,R)=(0,1,+1)$ subspace. Lattice size: $L=12$.  Left panel: 1D
chain.  Right panel: $3\times 4$ 2D lattice.}
\label{UnvsNPC}
\end{figure}

While the above results provide suggestive evidence in favor of using
entanglement as a diagnostic tool for integrability, independent
support is needed to rule out possible bias due to the special
relationship between multipartite entanglement and the c-basis, or to
the specific entanglement measure chosen.  A natural option for
circumventing such limitations is to invoke GE.  In particular, the
behavior of fermionic GE (as quantified by the
$P_{\mathfrak{u}(L)}$-purity defined in Eq.~(\ref{fermP})) as a
function of delocalization in the c-basis is depicted in
Fig.~\ref{UnvsNPC} for the same clean 1D and 2D models examined above.
Again, the observed delocalization dependence appears sensitively more
pronounced in the 2D chaotic case, as opposed to the 1D integrable
system.  It is intriguing to think that this behavior ultimately
reflects the fact that neighboring energy eigenstates possess common
features in the chaotic regime, as argued by Zelevinsky and coworkers
\cite{Zelevinsky1996} (see also \cite{Percival}).  Although these
findings reinforce the conjecture that, at least within the TBRE, a
new entanglement-based signature of chaos may be identified in this
way, validating this claim and its physical interpretation in more
generality requires a dedicated investigation which we plan to report
elsewhere.

\section{Conclusion and outlook}

We have provided a comprehensive quantitative analysis of spectral
properties, delocalization, and entanglement for the eigenvectors of
disordered spin-$1/2$ systems with Heisenberg interactions in one and
two spatial dimensions.  Disorder in the system has been introduced
through random applied magnetic fields, random interactions, or both.
Our main findings and conclusions may be summarized as follows.

(i) Although correspondence with non-integrability is well documented,
the interpretation of {\tt LSI} data is, as expected, non-trivial in
symmetry transition regions where invariant subspaces are partially
overlapping -- causing a tendency towards the Poisson distribution
regardless of whether integrability is approached or not.

(ii) Standard {\tt NPC} measures for state delocalization depend
entirely on which basis is chosen to represent the eigenvectors of the
Hamiltonian, thus they need not detect the transition from
integrability to chaos if the latter is not accompanied by a
significant change in the amount of delocalization. We have examined
the effect of basis choice in two distinct bases of limiting
integrable Hamiltonians as well as introduced a measure of relative,
disorder-dependent delocalization which is shown to successfully
detect quantum chaos in the presence of disorder.  We find that
delocalization, particularly in the computational basis, does not
achieve the level predicted by RMT even in chaotic regions for the
Heisenberg model.  Low connectivity and the localizing nature of the
Ising pairing are likely to play a role in accounting for such
disagreement.

(iii) In delocalized systems, {\tt NPC} and multipartite qubit
entanglement show a dependence on energy that resembles the one
observed for the density of states in the TBRE, reflecting a strong
correlation among the three quantities.  This complements findings on
the influence of LDOS properties to dynamical aspects of entanglement
in disordered qubit systems \cite{Montangero2006}.

(iv) In the case where only random Heisenberg interactions are
present, an interesting connection between level statistics, average
eigenvector entanglement and delocalization, and the presence of a
random singlet phase was uncovered in 1D.  This may warrant further
study, also in view of better elucidating what mutual relationships
(if any) exist between quantum criticality and the ability of a system
to sustain quantum chaos.

(v) Delocalization and multipartite entanglement both show a
substantial increase during the transition from a localized integrable
Hamiltonian to a chaotic one.  Similar to {\tt NPC} measures, both
entanglement and GE measures are unavoidably relative to a choice of
``local'' resources -- as captured by preferred subsystems or
observables.  However, comparison between two quantities sharing the
{\em same relative origin} -- {\tt NPC} and $P_n$ -- results in a
distinctive relationship which may serve as a quantum chaos indicator.
Physically, such a relationship suggests that both entanglement and
delocalization in a given local basis essentially capture the same
information about eigenvector structure.  Both the validity and
implications of this potential entanglement signature, as well as a
more substantial use of genuine GE, are points deserving additional
in-depth investigation.  In particular, a promising venue to explore
is the possibility that GE measures relative to appropriate observable
sets may allow to detect a transition to chaos starting from {\em any}
integrable model, irrespective of its local or non-local nature with
respect the the original operator language.

Finally, from a technical standpoint, the general method suggested in
\cite{pvn} and explicitly illustrated here for computing typical
entanglement properties of random pure states localized to a proper
subspace of states in Hilbert space is likely to find broader
applications within both quantum chaos and QIS.

\appendix
\section{Typical subspace entanglement: Expected values of block-purities}

Given the general expression of relative purity in Eq.~(\ref{hpur}),
each $P_n$ may be expressed as a sum of squared expectation values
over a normalized, orthonormal traceless basis $\{B_i\}$ for the
distinguished observable set $h$.  That is,
$$P_n(\ket{\psi}) = \kappa\sum_i \bra{\psi} B_i \ket{\psi}^2,$$
\noindent
where $\kappa$ is a normalization factor ensuring that
max$_{|\psi\rangle}\{P_n\} = 1$.  Thus, the average purity over any
ensemble of states is given by
$$\overline{P_n} (\ket{\psi}) = \kappa \sum_i \overline{\langle \psi
|B_i|\psi \rangle^2}.$$
\noindent
The method we shall follow to compute $\overline{P_n}$ will be based
on the following result (proved in \cite{pvn}):

{\em Let ${\cal H}$ be a finite-dimensional Hilbert space, with
dimension $N$, and let $A$ be any traceless real symmetric operator on
$\mathcal{H}$, normalized such that $\mbox{tr}(A^2) = N$.  Then for an
ensemble of pure states $|\psi\rangle$ with real components taken
uniformly with respect to the Haar measure on $O(N)$, the following
relationship holds:
$$\overline{\langle \psi | A |\psi \rangle^2}=\frac{2}{N+2}.$$}

For random states limited to a proper subspace,
$\mathcal{S}\subset\mathcal{H}$, additional care must be taken because
$A$ need not be normalized or traceless after projection onto
$\mathcal{S}$.  Let $N'$=dim(${\cal S}$) and let $\Pi$ denote the
projector onto $\mathcal{S}$.  If we let $\Pi A\Pi = \alpha A' +\beta
\mathbb{1}$, where $\mbox{tr}(A') = 0$ and $\mbox{tr}(A'^2) = N'$, it
follows that
$$\overline{\langle \Pi A\Pi \rangle^2}=
\frac{2\alpha^2}{N'+2}+\beta^2.$$
\noindent
The coefficients $\alpha$ and $\beta$ may be determined from
$\mbox{tr}(\Pi A\Pi)$ and $\mbox{tr}((\Pi A\Pi)^2)$,
$$\beta^2 = \frac{\mbox{tr}(\Pi A\Pi)^2}{N'^2},\;\;\; \alpha^2 =
\frac{\mbox{tr}((\Pi A\Pi)^2)}{N'}-\frac{\mbox{tr}(\Pi
A\Pi)^2}{N'^2},$$
\noindent
respectively. Therefore, in order to calculate $\overline{P_n}$, the
trace norm of both the projection $\Pi B_i\Pi$, and the projection
squared, $(\Pi B_i\Pi)^2$, of each basis operator onto $\mathcal{S}$
must be determined.

Since in our case each $P_n$ is an average over the purities of
$(L/n)$ $n$-qubit subsystems, it suffices to determine the average
subsystem purity of a single $n$-qubit subsystem.  A convenient
operator basis is provided by the set of $n$-qubit Pauli operators,
that is, all products of single-qubit Pauli operators and the
identity, $B_i = \sigma_{\alpha_1}^{(1)}
\otimes... \otimes\sigma_{\alpha_n}^{(n)}$, where $\alpha_k =0, x, y$,
or $z$. The choice of the identity, $\sigma_0$, acting simultaneously
on all qubits is excluded. It will be useful to consider the
representation of the Pauli operators as matrices expressed in the
standard c-basis.  The subspace ${\cal S}={\cal H}_0$ of states with
total $S_z=0$ angular momentum (no net magnetization) is relevant to
our study, $N'=N_0$.  The Pauli operators that have a non-vanishing
projection into $\mathcal{H}_0$ are:

(1) those consisting of only $\sigma_z$ and $\sigma_0$ operators; and

(2) those consisting of an even number of $\sigma_x$ and an even
number of $\sigma_y$ operators, with the remaining operators being any
combination of $\sigma_z$ and $\sigma_0$.

\vspace*{1mm}

First, consider case (1). Each such operator is diagonal in the
c-basis, hence it remains diagonal after projection into ${\cal H}_0$.
Because the eigenvalues of each such operator are $\pm1$,
$\mbox{tr}(\Pi B_i\Pi)$ may be determined by simply counting the
number of c-basis states spanning $\mathcal{H}_0$ that correspond with
each eigenvalue. For a Pauli string consisting of $m$ $\sigma_z$ and
$(n-m)$ $\sigma_0$ operators, the number of c-basis states with $k$
1's for the qubits acted on by the $\sigma_z$ operators is $m \choose
k$ $L-m \choose L/2-k$.  Thus, $$\mbox{tr}(\Pi B_i\Pi) = \sum_k (-1)^k
{m\choose k} {L-m\choose L/2-k}.$$
\noindent
If $m$ is odd, then the above quantity is $0$.

\vspace*{1mm}

Next, consider case (2). Note that the effect of the operator
$\sigma_x^{(i)}$ or $\sigma_y^{(i)}$ acting on a c-basis state is to
flip the i'th qubit from 0 to 1, or 1 to 0. Thus, for a c-basis state
to remain in $\mathcal{H}_0$ after the action of a Pauli operator, the
combined number of $\sigma_x$ and $\sigma_y$ operators must be even.
Additionally, since every state in the ensemble has only real
components in the c-basis, those Pauli operators that have an odd
number of $\sigma_y$ have zero expectation value for each state in the
ensemble. For a Pauli string with a total of $m$ $\sigma_x$ and
$\sigma_y$ operators, $m$ 0's and 1's will be flipped when the
operator acts on a c-basis state.  Thus, the qubits acted on must have
an equal number of 1's as 0's in order to remain in ${\mathcal{H}_0}$.
There are $m \choose m/2$ possible assignments of 0's and 1's for such
qubits. The remaining qubits (that is, those not acted upon by
$\sigma_x$ or $\sigma_y$ operators) must also have an equal number of
1's as 0's since the state is in ${\mathcal{H}_0}$. There are $L-m
\choose (L-m)/2$ possible assignments for such qubits. Therefore,
there are $m \choose m/2 $ $L-m \choose (L-m)/2$ matrix elements in
such a Pauli operator.  Since each matrix element is 1, it follows
that $$\mbox{tr}((\Pi B_i\Pi)^2) = {m \choose m/2} { L-m \choose
(L-m)/2}.$$
\noindent
Since $B_i$ acts non-trivially on every c-basis state, there are no
diagonal elements hence $\mbox{tr}(\Pi B_i\Pi) = 0$.

In order to properly normalize purity on an $n$-qubit subsystem,
consider the pure c-basis state $\ket{0,...,0}$. Only Pauli operators
consisting of $\sigma_z$ and $\sigma_0$ operators have non-vanishing
expectation values for this state, and each such expectation value is
1. There are $(2^n-1)$ such operators, therefore
$\kappa= {1}/{(2^n-1)}$ in this case.

With the above ingredients, $\overline{P_n}$ may be calculated by
determining how many Pauli basis states for an $n$-qubit subsystems
fall into each category described above in regard to trace norm and
squared trace norm upon projection onto ${\mathcal{H}_0}$.  We
consider various coarse-grained purities separately.

$\bullet \; P_1$: Only $\sigma_z^{(1)}$ has a non-vanishing
expectation value.  It has an odd number of $\sigma_z$ operators, thus
$\mbox{tr}((\Pi \sigma_z\Pi)^2) = N$ and $\mbox{tr}(\Pi \sigma_z\Pi) =
0$. Therefore,
$$\overline{P_1} = \frac{2}{N_0+2}.$$

$\bullet \; P_2$: The Pauli operators with non-vanishing expectation
values are: $\sigma_z^{(1)}$, $\sigma_z^{(2)}$, $\sigma_z^{(1)}
\sigma_z^{(2)}$, $\sigma_x^{(1)} \sigma_x^{(2)}$, and $\sigma_y^{(1)}
\sigma_y^{(2)}$. Thus,
$$\overline{P_2} = \frac{1}{3}\left \{\frac{2}{N_0+2}\left[3 -
\frac{\lambda_2^2}{N_0^2} + \frac{4}{N_0} {L-2 \choose (L-2)/2} \right] +
\frac{\lambda_2^2}{N_0^2}\right\},$$
\noindent
where for later purposes we let
$$\lambda_m = \sum_{k=0}^{k=m} (-1)^k {m\choose k} {L-m \choose
L/2-k}.$$

$\bullet \; P_3$: There are $n \choose m$ Pauli operators containing
$m$ $\sigma_z$ and $(n-m)$ $\sigma_0$ operators, and in addition there
are $2^{(n-2)}{n \choose 2} $ Pauli operators containing 2 $\sigma_x$
and $(n-2)$ $\sigma_0$ and $\sigma_z$ operators and the same number of
Pauli operators containing 2 $\sigma_y$ and $(n-2)$ $\sigma_0$ and
$\sigma_z$ operators.  This yields:
$$\overline{P_3}= \frac{1}{7}\left\{ \frac{2}{N_0+2}\left[
7-\frac{3\lambda_2^2}{N_0^2} + \frac{24}{N_0}{L-2 \choose (L-2)/2}\right]+
\frac{3\lambda_2^2}{N_0^2}\right\}.$$

$\bullet \; P_4$: All of the classes of Pauli operators listed for
$P_3$ must be considered for $P_4$ also, with the following additional
classes: the $2^{n-4}{n \choose 2,2,n-4}$ Pauli operators with exactly
2 $\sigma_x$ and 2 $\sigma_y$ operators; the $2^{n-4}{n \choose 4}$
Pauli operators with exactly 4 $\sigma_x$ operators; and the
$2^{n-4}{n \choose 4}$ Pauli operators with exactly 4 $\sigma_y$
operators.  This yields:
\begin{eqnarray*}
\overline{P_4}& = &  \frac{1}{15}\left\{ \frac{2}{N_0+2}\left[15
    -\frac{6\lambda_2^2}{N_0^2}-\frac{\lambda_4^2}{N_0^2}
+ \frac{48}{N_0}{L-2 \choose
(L-2)/2} \right.\right.
\\
&+& \left. \left. \frac{8}{N_0}{L-4 \choose (L-4)/2} \right]+
\frac{6\lambda_2^2}{N_0^2}+\frac{\lambda_4^2}{N_0^2}\right\}.
\end{eqnarray*}

$\bullet \; P_6$: The additional classes of Pauli operators to
contribute to $P_6$ are: the $2^{n-6} {n\choose 4,2,n-6}$ Pauli
operators with exactly 4$ \sigma_x$ and 2 $\sigma_y$ operators; the
$2^{n-6} {n\choose 2,4,n-6}$ Pauli operators with exactly 2 $\sigma_x$
and 4 $\sigma_y$ operators; the $2^{n-6}{n\choose6}$ Pauli operators
with exactly 6 $\sigma_x$ operators; and the $2^{n-6}{n\choose6}$
Pauli operators with exactly 6 $\sigma_y$ operators.  This yields:
\begin{eqnarray*}
\overline{P_6}&=& \frac{1}{63}\left \{ \frac{2}{N_0+2}\left[ 63
  -\frac{15\lambda_2^2}{N_0^2} -\frac{15\lambda_4^2}{N_0^2}
  -\frac{\lambda_6^2}{N_0^2} \right. \right. \\ &+& \frac{480}{N_0}{L-2
    \choose (L-2)/2}+ \frac{480}{N_0}{L-4 \choose (L-4)/2}\\ &+ &
  \left. \left.\frac{32}{N_0}{L-6 \choose (L-6)/2} \right]
+\frac{15\lambda_2^2}{N_0^2}+\frac{15\lambda_4^2}{N_0^2}
+\frac{\lambda_6^2}{N_0^2}\right\}.
\end{eqnarray*}
Letting $L=12, N_0=924$ in the above formulas leads to the $n$-local
purity values quoted in Table I.

\begin{acknowledgments}

It is a pleasure to thank Viatcheslav V. Dobrovitski, Simone
Montangero, and Yaakov S. Weinstein for valuable discussions and
feedback.  D. S. contribution to this work was initiated as part of an
{\em Undergraduate Research Experience in Quantum Information Science}
during Summer 2005, when he was a junior at SUNY Fredonia, NY.
W. G. B. gratefully acknowledges partial support from Constance and
Walter Burke through their Special Projects Fund in Quantum
Information Science, and from a GAANN Fellowship.

\end{acknowledgments}


\begin{thebibliography}{95}
\bibitem[{rem()}]{remark1}
\bibinfo{note}{We verified that using Gaussian-distributed random numbers in
  place of uniform random numbers produced no significant qualitative
  difference.}
\bibitem[{tec()}]{technote}
\bibinfo{note}{Technically, for TBRE, recovery of GOE statistics requires
  unfolding the spectrum of each member individually, as implemented in our
  numerical procedure.}
\bibitem[{Rem({\natexlab{b}})}]{Remark0}
\bibinfo{note}{As a particular feature of the $S_z=0$-subspace of ${\cal H}_0$,
  note that it is also invariant under the discrete transformation $Z_2 =
  \otimes_{i=1}^L \sigma_x^{(i)}$, which is associated to a global $\pi$
  rotation about the $x$-axis. However, since the value of the total angular
  momentum determines the irreducible representation according to which each
  state transforms under rotations, such a $Z_2$-symmetry is not independent of
  $S^2$, and consequently does not induce additional invariant subspaces.}
\bibitem[{com({\natexlab{b}})}]{comment}
\bibinfo{note}{Note that the transition to non-integrability occurs at a larger
  value than in the other disorder setting, e.g. $\delta J/\delta \varepsilon
  \sim 10$ in 1D. The reason for this shift is simply due to the fact that,
  given the parametrization used, the range of the interaction is chosen, but
  the ratio between the {\em average} interaction strength and the on-site
  disorder strength should be considered for a fair comparison.}
\bibitem[{Ham()}]{Hamming}
\bibinfo{note}{Given two binary strings of equal length, the Hamming distance
  counts the number of substitutions needed to change one into the other.
  Physically, this has a simple interpretation in terms of spin-flip
  operations.}
\end{thebibliography}

\end{document}